# Chasing the eternal sun: Does a global super grid favor the deployment of solar power?

Xiaoming Kan[a,*]   Fredrik Hedenus[a]   Lina Reichenberg[a]

[a]Department of Space, Earth and Environment, Chalmers University of Technology, 41296 Gothenburg, Sweden

## Abstract

The *One Sun One World One Grid* (OSOWOG) initiative advocates the development of a global *Super grid* for sharing renewable energy, especially solar energy. This study evaluates the economic benefits of such a *Super grid*, which connects six large regions spanning from Australia to the US, utilizing a detailed energy system optimization model and considering heterogeneous discount rates among countries. Integrating the six regions into a *Super grid* reduces the electricity system cost by 3.8% compared to isolating them. In contrast, grid expansion within each region reduces the electricity system cost by 12% on average. The economic benefits of the OSOWOG initiative's global *Super grid* expansion seem to be rather limited. Moreover, the allowance for a *Super grid* consistently results in decreased investments in solar power, indicating that it is not an effective strategy for enhancing the deployment of solar power, even when transmission grids covering 18 time zones are available.

## 1. Introduction

The past decade has witnessed substantial cost reductions and rapid deployment of variable renewable energy (VRE) technologies such as wind power and solar photovoltaic (PV) [1]. In cost-optimal scenarios for future low-carbon energy systems, wind and solar power often serve as the cornerstone of the energy supply [2-8]. Solar energy, although abundant and widely available, is limited to the daytime and subject to weather conditions [9]. However, from a global perspective, the sun never sets, as half of the earth is bathed in sunshine at any given time. The concept of continuously exploiting the ceaseless solar radiation involves the construction of an intercontinental transmission network that connects different time zones and facilitates the trading of solar energy across time zones. In line with this idea, the Prime Ministers of India and the UK jointly launched the *Green Grids - One Sun One World One Grid* (OSOWOG) initiative at the COP26 UN Climate Change Conference in Glasgow. The objective of this initiative is to facilitate the development of a *Super grid* that would cover the entire globe, so as to promote the integration of solar energy and transmit clean energy globally at all times [10-12]. The OSOWOG initiative is planned to have three phases [11, 13]. In the first phase, the Indian electricity grid will be connected to the grids in South and Southeast Asia and the Middle East. In the second phase, this grid will be connected to African regions with abundant renewable energy resources. Finally, the third phase will complete a global, interconnected network that can be accessed by all countries. An expanded grid serves two main functions for a renewable energy system. First, it enables *resource tapping*, allowing areas with cheap and abundant solar and wind resources to export electricity to regions with high electricity demands. Second, it facilitates *variation management*, addressing the temporal variability of power production from VRE resources through spatial connections to regions with compensating generation patterns. In this study, we assess the potential benefits of constructing the global *Super grid* proposed by the OSOWOG initiative. We categorize the transmission grids within a large region (e.g., South Asia) or a continent as the *Regional grid*, and define the transmission grids connecting multiple continents as the *Super grid*. These terms will be consistently employed throughout the remainder of this study.

---

[*] Correspondence: kanx@chalmers.se

Several studies [3, 14-18] investigated the benefits of integrating two or more continents with transmission grids for a future renewable electricity system. These studies showed that connecting multiple continents, in contrast to isolating them, can reduce electricity system costs by up to 5% for Eurasia [16], 1.6% for the Americas [14], 1.3% for Eurasia, the Middle East and North Africa [15], and 2% for the entire world [3]. In addition, Guo et al. [18] explored the decarbonization pathway for the entire world, and suggested that introducing a global *Super grid* can reduce the electricity system cost by up to 2%. In stark contrast, Prol et al. [17] reported that global electricity trade, achieved by combining complementary seasonal and diurnal cycles of solar power production, could result in a 74% reduction in electricity system cost for most parts of the world.

In terms of variation management, it is well-established in the literature that extending transmission grids is a cost-effective strategy for managing the variability of wind power [5, 6, 19, 20]. For solar power, some studies suggested that a *Super grid* connecting multiple time zones could help to alleviate the intermittency of solar power by transmitting solar energy to regions that experience nighttime or winter seasons [17, 21-25]. Moreover, two other studies argued that linking regions in different time zones could completely eliminate the need for dispatchable energy resources or storage in a solar power-based system [26, 27].

Based on the literature, it seems that a *Super grid* connecting several different continents can be beneficial or cost-effective in certain cases. Studies that conducted oversimplified analyses, such as excluding wind power and assuming zero costs for transmission grids [17], typically found a *Super grid* to be more attractive, as compared to more comprehensive energy system analyses [3, 14-16, 18]. As for the spatial smoothing effect of grid connections for solar power, previous research generally focused on assessing the physical feasibility of combining solar power generation patterns across different time zones [21-27]. However, it remains unknown as to whether such a combination is an effective strategy for deploying solar power within a comprehensive energy system. Furthermore, we observe that in previous intercontinental energy system modeling studies [3, 14-18], a uniform discount rate was assumed globally, despite the substantial variation in discount rates between countries [28, 29]. It is important to note that the discount rate strongly influences the cost-competitiveness of capital-intensive energy technologies [28-31].

The contributions of this study are threefold. First, we evaluate specifically the impacts of a global *Super grid*, as proposed by the OSOWOG initiative, on the cost and configuration of a future renewable electricity system. Second, in contrast to previous studies that focused solely on the a *priori* benefits of combining compensatory solar power generation patterns across different time zones [21-27], we examine whether extending transmission grids across up to 18 time zones could promote solar power deployment within a complex energy system that incorporates various energy technologies. Third, we analyze how the heterogeneity of discount rates across countries affects the expansion of a global *Super grid* and the associated benefits.

## 2. Methods

We use a techno-economic cost optimization model with hourly time resolution to model six interconnected sunny regions: Australia (AU), South Asia (SA), the Middle East and North Africa (MENA), Central and South Europe (CSE), South America (SAM), and Central and North America (CNA) (Fig. 1). The six regions include all the member countries[1] of the OSOWOG initiative, and each region is divided into several subregions. In total, this study covers 48 subregions spanning 18 time zones (Fig. 1). The benefits of the global *Super grid*, as proposed by the OSOWOG initiative, are assessed for a renewable electricity system[2] in Year 2050, considering various assumptions with respect to technology costs, heterogeneous discount rates among countries, the availability of nuclear power, hydrogen production, and uncertainty related to future electricity demands.

Our evaluation focuses on the electricity system cost and the electricity supply mix under three distinct levels of transmission connection: 1) *Isolation* – the subregions within each region are isolated from each other without transmission connections; 2) *Regional grid* – transmission expansion is permitted within each region to connect all the subregions; 3) *Super grid* – transmission expansion is allowed to connect all six regions.

---

[1]The member countries and the Steering Committee of the OSOWOG initiative include Australia, India, the UK, France and the US.
[2]This system is primarily dominated by wind and solar power, complemented mainly by hydropower, biogas power plants and battery storage.

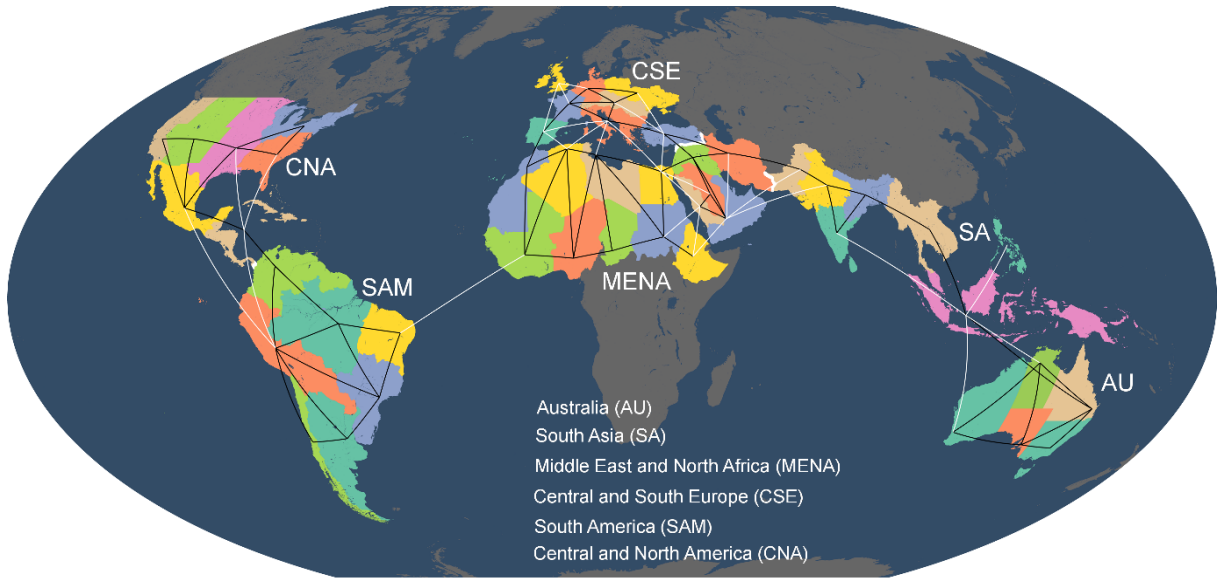

Fig. 1 The modeled interconnected regions, the subregions inside each region, and the potential transmission network topology connecting all the subregions. See Figs. S1–S6 for details of each individual region.

**2.1 Energy system model**

In this study, we model six interconnected regions with the open data and open source Supergrid model [32], which is a greenfield capacity expansion model with an hourly time resolution. The model optimizes investment and dispatch of the electricity sector with an overnight approach. The exception is hydropower, where existing hydropower plants are assumed to be still in operation in Year 2050, and their capacity is assumed to remain at the current level due to environmental regulations. In terms of the $CO_2$ emissions target, we assume a near-zero emissions system with a global $CO_2$ emission cap of 1 $gCO_2$ per kWh of electricity demand. The model is written in the Julia programming language using the JuMP optimization package. The cost assumptions and key parameters for technologies are summarized in Table 1. For a more detailed description of the model, see Mattsson et al. [32]. The model-specific code and input data can be found at this link: https://github.com/xiaomingk/Supergrid.

**Table 1** Cost data and technical parameters.

| Technology | Investment cost [$/kW] | Variable O&M costs [$/MWh] | Fixed O&M costs [$/kW/yr] | Fuel costs [$/MWh fuel] | Lifetime [years] | Efficiency/ Round-trip efficiency |
|---|---|---|---|---|---|---|
| Natural gas OCGT | 500 | 1 | 10 | 22 | 30 | 0.35 |
| Natural gas CCGT | 800 | 1 | 16 | 22 | 30 | 0.6 |
| Coal | 1600 | 2 | 48 | 11 | 40 | 0.45 |
| Nuclear | 5000 | 3.5 | 112 | 2.6 | 40 | 0.33 |
| Biogas OCGT | 500 | 1 | 10 | 37 | 30 | 0.35 |
| Biogas CCGT | 800 | 1 | 16 | 37 | 30 | 0.6 |
| Onshore wind [a] | 825 | 0 | 33 | n/a | 25 | n/a |
| Offshore wind [a] | 1500 | 0 | 55 | n/a | 25 | n/a |
| Solar PV [b] | 323 | 0 | 8 | n/a | 25 | n/a |
| Solar Rooftop [b] | 423 | 0 | 5.8 | n/a | 25 | n/a |
| CSP | 3746 | 2.9 | 56 | n/a | 30 | n/a |
| Electrolyzer | 250 | 0 | 5 | n/a | 25 | 0.66 |
| Hydrogen storage | 11 $/kWh | 0 | 0 | n/a | 20 | n/a |
| Fuel cell | 800 | 0 | 40 | n/a | 10 | 0.5 |
| Hydro | 300[c] | 0 | 25 | n/a | 80 | 1 |
| Onshore Transmission[d] | 400 $/MW/km | 0 | 8 $/MWkm | n/a | 40 | 0.016 loss per 1000 km |
| Offshore Transmission[e] | 470 $/MW/km | 0 | 1.65 $/MWkm | n/a | 40 | 0.016 loss per 1000 km |
| Converter[d] | 150 | 0 | 3.6 | n/a | 40 | 0.986 |
| Battery[f] | 116 $/kWh | 0 | 1.5 $/kWh | n/a | 15 | 0.9 |

[a]IRENA [33]

[b]IRENA [34]

[c]Steffen [35], this cost refers to the expenses associated with the replacement of old mechanical and electrical machinery.

[d]Hagspiel et al. [36]

[e]Purvins et al. [37]

[f]Cole et al. [38]

O&M, operation and maintenance; OCGT, open-cycle gas turbine; CCGT, combined-cycle gas turbine; CSP, concentrating solar power.

The main scenarios explored in this study are outlined in Table 2. The economic benefits of a *Super grid* are examined across a broad spectrum of cost assumptions for transmission grid and nuclear power. Onshore transmission gird costs are in the range of 150–950 $/MW/km, offshore transmission grid costs 200–1000 $/MW/km, and nuclear power costs 2000–8000 $/kW. For the sensitivity analysis, onshore wind power investment costs are set at 650 ('Low'), 825 ('Mid'), 1000 ('High') and 1715 ('Extremely

high') $/kW, solar PV investment costs are set at 165 ('Low'), 323 ('Mid') and 481 ('High') $/kW, battery storage investment costs are set at 76 ('Low'), 116 ('Mid'), 156 ('High') and 385 ('Extremely high') $/kWh, and concentrating solar power (CSP) investment costs are set at 3746 ('Default') and 6500 ('Extremely high') $/kW. For onshore wind power, CSP and battery storage, the 'Extremely high' costs are identical to the present values.

**Table 2** Scenarios of this study.

| Scenario | Wind cost | Solar cost | Storage cost | CSP cost | Nuclear availability | Nuclear cost | Demand | Discount rate |
|---|---|---|---|---|---|---|---|---|
| Base | Mid | Mid | Mid | Default | No | - | Default | Country-specific |
| Uniform discount rate | Mid | Mid | Mid | Default | No | - | Default | Uniform |
| Plus East Asia[a] | Mid | Mid | Mid | Default | No | - | Default | Country-specific |
| Low nuclear | Mid | Mid | Mid | Default | Yes | 2000 $/kW | Default | Country-specific |
| Double demand | Mid | Mid | Mid | Default | No | - | Double | Country-specific |
| Hydrogen | Mid | Mid | Mid | Default | No | - | Plus hydrogen demand | Country-specific |
| Solar friendly | Extremely high | Low | Extremely high | Extremely high | No | - | Default | Country-specific |

[a]In this scenario, East Asia is added as an additional region to the six regions included in the **Base** scenario.

## 2.2. Transmission assumptions

In this study, all the six regions are divided into several subregions (see Figs. 1, S1–S6), and we assume that the subregions can be interconnected via high-voltage direct current transmission grids. The electricity trade is treated as a simple transport problem [20, 39], and all the subregions in the model are assumed as "copper plates" without intraregional transmission constraints. Transmission costs are estimated based on whether the connection is entirely overland or partially marine, and on the length of the transmission line, which is measured as the distance between the population centers of the individual subregions [32].

## 2.3. Wind, solar and hydro data

An important parameter for estimating the renewable energy supply potential is how densely wind and solar power can be installed in the landscape. For this study, we first exclude areas unsuitable for large-scale wind and solar power plants. A fraction of the remaining land is then utilized as the available land for wind and solar power installations. Specifically, protected areas are excluded from the installation of wind and solar. For the remaining areas: utility-scale solar units may be placed on all land types except forests; solar rooftop may be placed in urban areas; onshore wind power may be placed on all land types except densely populated areas (population density >500 people per km$^2$); and offshore wind power may be placed on the seabed at a depth of up to 60 m. We assume that 5% of the suitable area can be used for solar PV, rooftop PV and CSP, and that 10% can be used for onshore and offshore wind power (see Table 3). For a detailed analysis of onshore wind power deployment, see Hedenus et al. [40]. The capacity factors for wind and solar are computed using the ERA5 reanalysis data (hourly wind speed, direct and diffuse solar insolation) [41] and the annual average wind speed from the Global Wind Atlas [42] for Year 2018. In the case of solar PV, we assume a fixed-latitude-tilted PV technology. The capacity factor for wind power is calculated using the power curve of a typical wind farm equipped with Vestas V112-3.075-MW wind turbines. To represent accurately the capacity factors for wind and solar power, we categorize these technologies into five classes based on resource quality [32]. Additionally, we model CSP with 10 hours of thermal storage.

The existing hydropower capacity, reservoir size and monthly inflow are obtained from previous studies [43-45]. For those cases where data for certain regions are unavailable, we adopt a conservative assumption, setting the reservoir capacity as equivalent to 6 weeks of peak hydropower production. For regions where it is challenging to distinguish between reservoir and run-of-river plants due to data limitations, we assume that a minimum of 40% of hourly water inflow must be utilized for electricity generation to constrain the flexibility offered by hydropower.

**Table 3** Assumptions regarding the capacity limits for wind and solar PV.

|  | **Solar PV** | **Solar Rooftop** | **CSP** | **Onshore wind** | **Offshore wind** |
| --- | --- | --- | --- | --- | --- |
| Density [W/m$^2$]$^a$ | 45 | 45 | 35 | 5 | 5 |
| Available land [%] | 5% | 5% | 5% | 10% | 10% |

$^a$The term 'Density' refers to the capacity assumed to be installed per unit area for a typical solar or wind farm.

**2.4. Demand data**

The future electricity demand is projected with the open data, open source GlobalEnergyGIS package [32]. We first estimate the annual electricity consumption for each region in Year 2050 based on the annual demand in Year 2016 [46] and the regional demand growth between 2016 and 2050 in the Shared Socioeconomic Pathway 2 scenario [47]. We then estimate the hourly demand profile based on a machine learning approach, which adopts the historical demand profiles for 44 countries as input to a gradient boosting regression model [48]. The regression model takes into account the calendar effects (e.g., hour of day, weekday and weekend), temperature (e.g., hourly temperature in the most populated areas of each region), and economic indicators (e.g., local GDP per capita). Finally, the hourly demand series is scaled to match the annual electricity demand for each region in Year 2050. Regarding hydrogen demand, we assume that the annual demand for hydrogen is equivalent to half of the annual electricity demand. This assumption aligns with the scale of the projected hydrogen demand for Year 2050, as outlined in the European Commission's long-term strategic vision [49].

**2.5. Discount rate**

The fixed investment costs of renewable energy technologies are usually represented with an overnight capital cost (OCC), which is depreciated over the economic lifetime using a weighted average cost of capital (WACC) [50]. Both the OCC and WACC are project-specific and can vary by region and over time. Typically, OCC covers the costs for materials, equipment and labor [51]. WACC incorporates the financing structure of a project, which includes the costs of equity and debt, as well as government support, such as subsidies [52]. The constituent components of WACC and OCC can vary from project to project, across technologies and industries, and are heavily influenced by national and local priorities.

We recognize that it is almost impossible to estimate average values for OCC and WACC for all of the countries in the world based on a bottom-up approach that accounts for all the relevant items analyzed above. Instead, we focus on the regional risk premium that can drive the capital investment costs of a project and the cost of servicing such investments [53]. Specifically, we assume a uniform capital cost for all projects and discount the capital cost over the lifetime with country-specific discount rates that incorporate risk premium estimates from Damodaran [53]. The country-specific discount rate is calculated by adding the risk premium to a "risk-free" baseline discount rate (5%). The data for the risk premiums are available for most countries in the world. In cases where specific data are unavailable for certain countries, the average value derived from the neighboring countries is assigned. For each subregion covered in this study, the discount rate is determined by averaging the country-specific discount rates of the countries included in that subregion. Thereafter, the discount rate for the transmission line is decided by the node with the higher discount rate.

## 3. Results

**3.1 Cost savings attributed to allowing for a *Super grid***

We explore the potential benefits of transmission grid expansion, as proposed by the OSOWOG initiative, by assessing the electricity system cost and the electricity supply mix under three distinct levels of transmission connection: *Isolation*, *Regional grid* and *Super grid*. As shown in Fig. 2, allowing for transmission grid expansion inside each region consistently reduces the electricity system cost, as compared to isolating the subregions. The average electricity system cost reduction due to *Regional grid* expansion is 12% (14% for Australia, 7% for South Asia, 14% for the Middle East & North Africa, 16% for Central & South Europe, 13% for South America and 13% for Central & North America). In contrast, the reduction in system costs attributed to the *Super grid*, i.e., enabling transmission grid expansion between all six regions, is only 3.8%. These findings indicate a significantly stronger impact on the system cost of developing *Regional grid*s compared to integrating continents into a *Super grid*. Thus, the economic benefits of a global *Super grid*, as suggested by OSOWOG, are likely to be rather limited in a renewables-based system.

We also explore the potential benefits of grid connections between India and its neighboring countries, following the plans outlined in the different phases of the OSOWOG initiative. Integrating the Indian grid with countries in Southeast Asia could result in a 7% reduction in the overall electricity system cost (Fig. S7), while connecting India with countries in the Middle East confers a 12% reduction in the electricity system cost (Fig. S8). In comparison, further integrating the *Regional grid* in Asia to Africa results in only a 1% reduction in the electricity system cost (Fig. S9). It appears that linking India to its neighboring countries may offer more substantial economic benefits than extending the *Regional grid* in Asia to Africa.

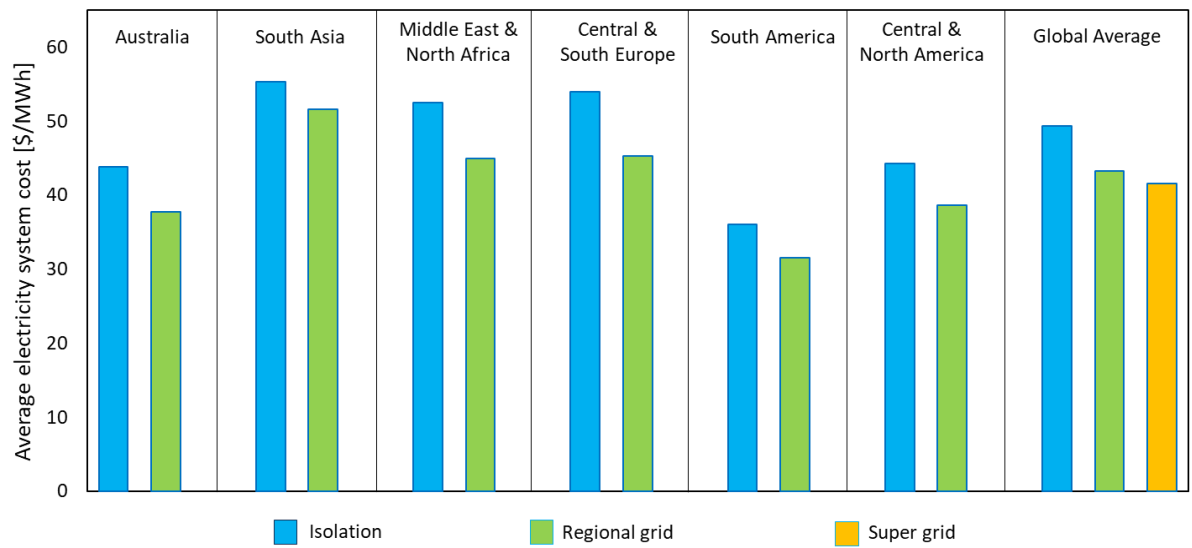

Fig. 2 Average system cost for a renewable electricity system at three different levels of transmission connection: *Isolation*, *Regional grid* and *Super grid*.

### 3.2 Transmission grid expansion and solar power deployment

Investment in intercontinental transmission grids occurs mainly between MENA and CSE (212 GW) and between AU and SA (108 GW)[3] (Fig. 3). Both CSE and SA are characterized by a high electricity demand, while MENA and AU are endowed with substantial high-quality renewable energy resources. In a renewable future, building a *Super grid* allows these regions to reap mutual benefits, where electricity is traded from regions with high-quality renewable resources to regions with a high demand for electricity. The creation of producer/consumer-regions that arise from the availability of intercontinental transmission grids is depicted in Fig. 3, which shows an evident increase in electricity production in MENA and AU when the six regions are integrated, whereas there is a marked decline in the electricity generation in CSE and SA.

---

[3] The transmission capacity connecting MENA and CSE amounts to 30% of the peak demand in CSE, while the transmission capacity linking AU and SA equals 14% of the peak demand in SA.

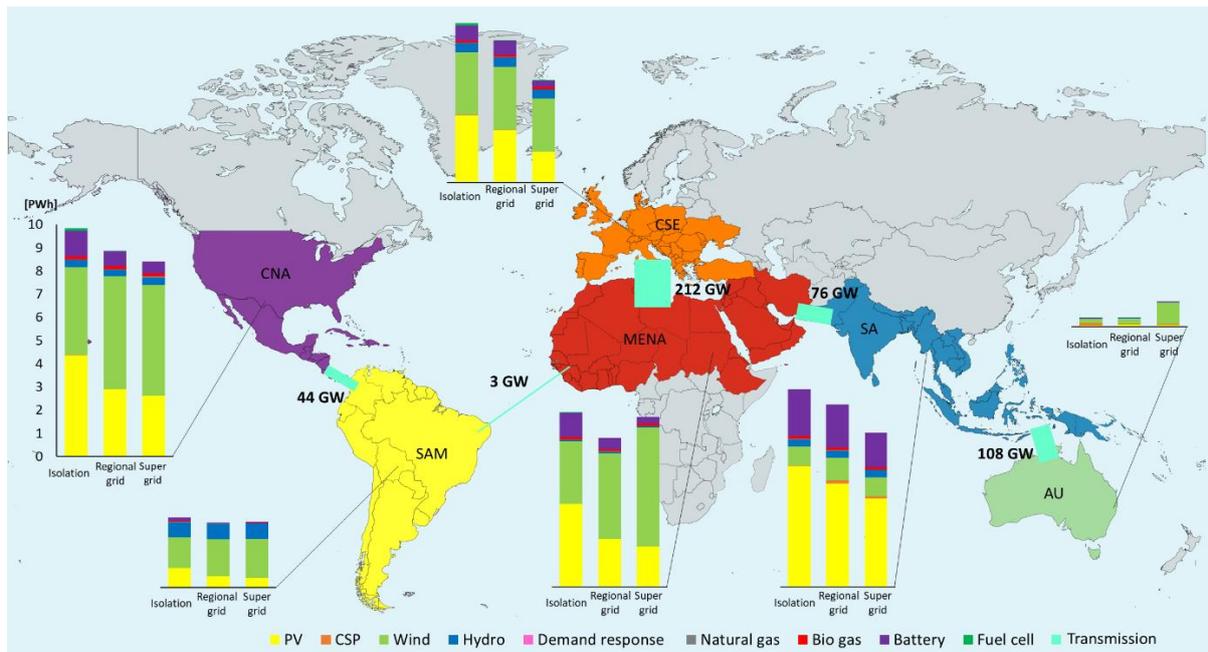

Fig. 3 Optimal transmission capacities between regions when the six regions are integrated, and the electricity supply mix for each region at three different levels of transmission connection: *Isolation*, *Regional grid* and *Super grid*.

The OSOWOG initiative suggests developing an intercontinental *Super grid* to advance the deployment of solar PV. Our results show the opposite effect: the consequence of introducing a *Super grid* (connecting 18 time zones) is that there is less solar PV in the electricity supply mix compared to *Regional grid* expansion (Fig. 4). In other words, solar PV is less cost-effective in the world envisioned by the OSOWOG initiative. This phenomenon is evident not only in the overall electricity supply mix for all regions, but also in the electricity supply mix for MENA (see Fig. 3). MENA has abundant high-quality solar resources, yet building a *Super grid* to encompass this region does not enhance the integration of solar PV. To reveal more clearly the potential benefits of a *Super grid* for solar power development, we investigate one extreme scenario (**Solar friendly** scenario) with a low cost for solar power, and extremely high costs (same as the present costs) for wind power, CSP and battery storage. A high cost for wind power means that solar power is more competitive. High costs for CSP and battery storage entails that domestic variation management is expensive for solar power, which potentially favors the spatial smoothing of the diurnal variation for solar power over a broad span of time zones. However, even with such a favorable cost configuration, no expansion of solar PV is observed in the optimal electricity supply mix when the *Super grid* option is enabled (see Fig. 4). This result confirms that transmission grid and trade do not represent an effective tool for deploying solar power, even in the presence of transmission grids covering 18 time zones.

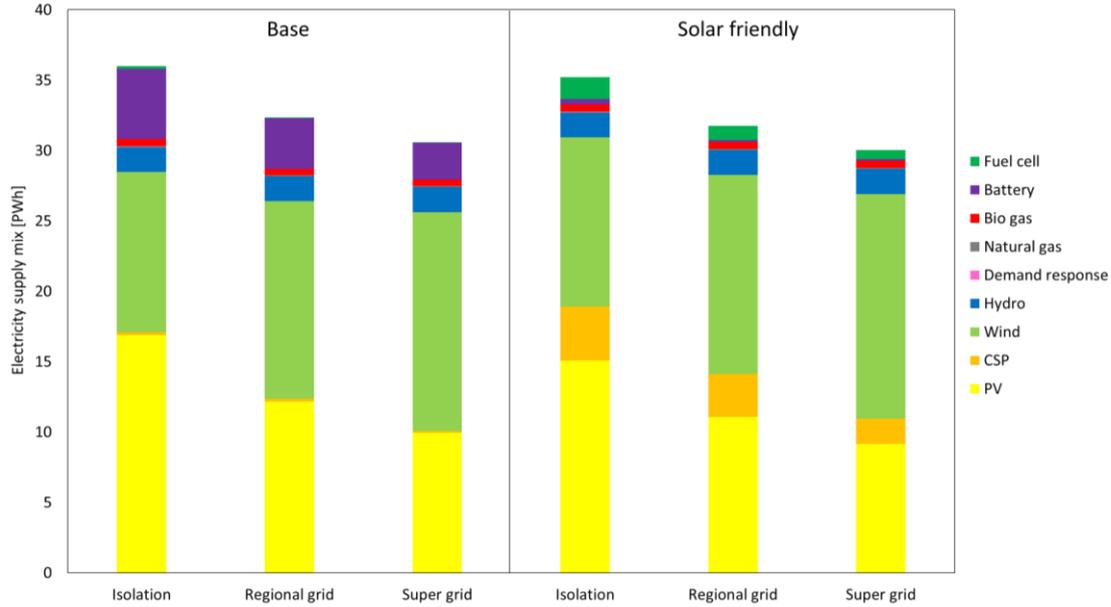

Fig. 4 Electricity supply mixes for the **Base** scenario and **Solar friendly** scenario in which the cost assumptions for wind power, CSP and battery storage are kept at the current values. For a description of all the scenarios included in this study, see Table 2 in the *Methods* section.

Notably, as the share of solar PV in the electricity supply mix decreases, there is also a reduction of battery storage in the optimal electricity supply mix (**Base** scenario, Fig. 4). The decrease in battery storage is not surprising, as a battery is primarily a short-term storage technology to complement the diurnal variation of solar PV [6, 20]. In comparison, connecting the six regions consistently increases the share of wind power in the optimal electricity supply mix. The increased deployment of wind power due to grid integration is consistent with the findings of national and continent-wide energy system studies [5, 6, 19, 20].

### 3.3 Impacts of country-specific discount rates

Our main results are based on country-specific discount rates, reflecting a future in which the discount rate for each country remains the same as today. However, in an optimistic future characterized by sustained political stability and consistent economic growth, the diverse discount rates across countries may gradually converge toward a common, low-risk mean value. In such a future (**Uniform discount rate** scenario), the economic benefit of a global *Super grid* increases to 6.2%, as compared to the **Base** scenario with a system cost reduction of 3.8%. The lower level of benefit observed when applying country-specific discount rates is primarily attributed to the shift in investments in renewable energy towards low-risk regions (e.g., Europe) that have renewable resources of lower quality (Figs. S10 and S11). This shift weakens the resource-tapping function of a *Super grid*. A typical example relates to MENA and Europe. With a uniform discount rate, Europe is heavily dependent on importing renewable energy from MENA (see Fig. S11a). Notably, African countries exhibit surplus electricity generation due to competitive renewable resources. In contrast, when country-specific discount rates are applied, the trade in electricity from MENA to Europe experiences a significant reduction (see Fig. S11b). As for intercontinental transmission connections, the transmission capacity between MENA and Europe decreases from 356 GW to 212 GW when accounting for the heterogeneity of discount rates. Given these outcomes, using Africa to tap the abundant renewable energy resources for Europe would require favorable socio-political developments in Africa.

**3.4 Cost savings and solar power deployment in a wide range of scenarios**

To assess whether an even larger geographic scope could enhance the benefits of an intercontinental *Super grid*, we include East Asia in our analysis (**Plus East Asia** scenario). Extending the *Super grid* to the substantial electricity demand center in East Asia does not amplify its benefits; the overall benefit of the *Super grid* is 3.4%, closely aligning with the benefit (3.8%) of connecting the original six regions (Fig. 5).

If there are alternative ways to generate low-carbon electricity, such as nuclear power, these may diminish the advantages of a *Super grid* by weakening the dependence on renewable resource sharing and reducing the demand for variation management. We simulate this by including nuclear power in the analysis. The economic gain associated with allowing for a *Super grid* is less than 1.4% when nuclear power is cheap (**Low nuclear** scenario) (Fig. 5, Fig. S12).

To examine the impacts of extensive electrification and electricity-derived fuel production on the benefit of a *Super grid*, we consider scenarios that involve doubling the electricity demand (**Double demand** scenario) and integrating hydrogen production (**Hydrogen** scenario). The economic benefit of the *Super grid* increases from 3.8% to 5.1% with a doubling of the electricity demand (Fig. 5). In a renewable future, countries that lack abundant high-quality renewable resources may need to import electricity, particularly with increasing demand for electricity. Large-scale hydrogen production significantly reduces the benefit of a *Super grid* to only 1.4% (Fig. 5). The flexibility provided by hydrogen production serves as an alternative variation management strategy, which weakens the impact of transmission grid expansion.

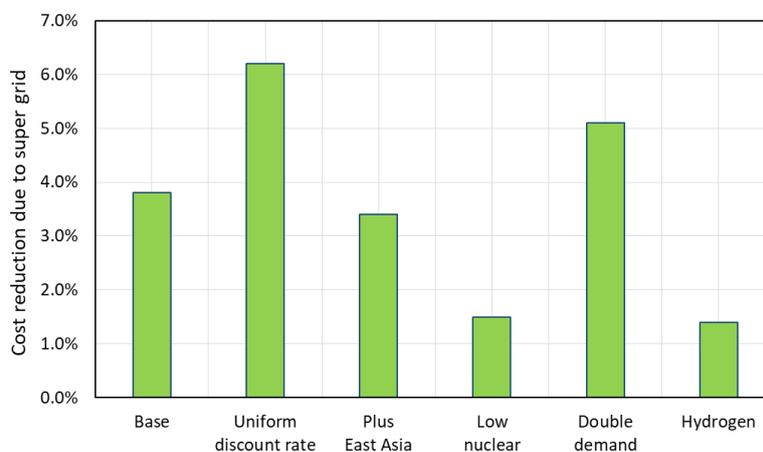

Fig. 5 Electricity system cost reductions linked to allowing for a *Super grid* for various scenarios.

The transmission cost is a crucial factor for the development of an intercontinental *Super grid*. Therefore, we evaluate the benefits of a *Super grid* under different transmission cost assumptions. The reduction in electricity system cost is greater than 5.4% if either onshore or offshore transmission grid is exceptionally cheap, with the most substantial cost reduction being 6.7% (Fig. 6). In general, a high cost for transmission grid diminishes the benefit of a *Super grid* to less than 1.6%. To understand further the broader conditions that may affect the cost-effectiveness of a *Super grid*, we evaluate the benefits of a *Super grid* under various cost assumptions for wind and solar power and battery storage. Across the broad range of cost assumptions, the economic benefit of connecting the six regions via a *Super grid* is less than 6.1% (Fig. S13). The largest cost reduction is achieved when solar PV and battery storage costs are high, and the cost of wind power is low. Such a cost combination favors investments in wind power, and a *Super grid* allows for resource tapping and variation management for the more competitive wind power over a larger geographic area.

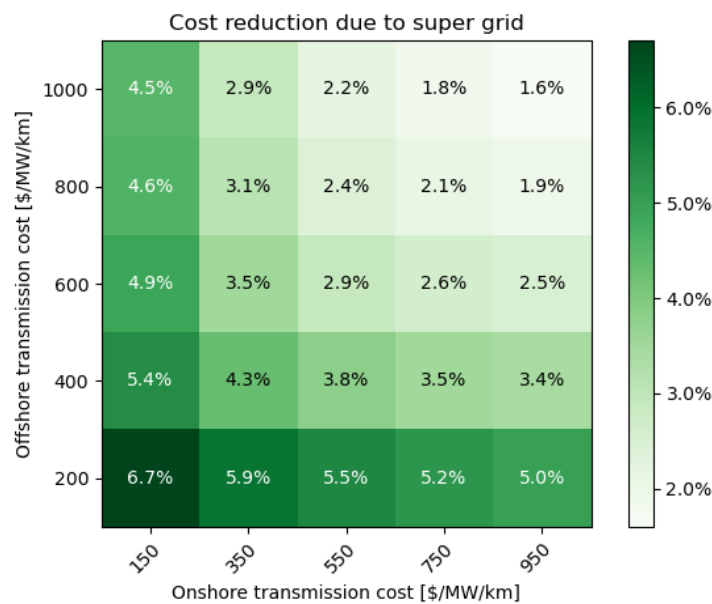

Fig. 6 Electricity system cost reductions linked to allowing for a *Super grid* for different cost assumptions for the transmission grid.

For all the scenarios investigated above, integrating the six regions into a *Super grid* does not increase the deployment of solar power (Figs. S14-S18). These outcomes appear to contradict the findings of previous studies that explored the physical feasibility of harmonizing solar power generation patterns across diverse time zones to manage variations in solar power production [21-27]. To delve deeper into this disparity, we conduct additional experiments to pinpoint the circumstances under which a *Super grid* could indeed bolster solar power deployment. Our findings indicate that an evident uptick in solar power deployment occurs only when the transmission cost falls below 50 $/MW/km (Fig. 7). However, this cost is deemed unrealistically low. To provide context, this cost represents less than half of the estimated future transmission grid cost in China [54], a country that is renowned for its capability to construct extensive transmission networks at low cost. From the physical standpoint, the notion of

transmitting solar energy from regions with peak solar power production, such as Morocco, to areas that are experiencing evening peak demand, such as India, presents an appealing solution for managing the diurnal variation in solar power production. However, facilitating such an energy transfer necessitates substantial expansion of extremely long-distance transmission grids connecting India and Morocco. While this approach may a *priori* seem to be an effective way to manage the intermittency of solar power production, these advantages are ultimately offset by the substantial costs of the transmission grids. This explains why the increased share of solar energy facilitated by a *Super grid* only appears when transmission grid costs are unrealistically low.

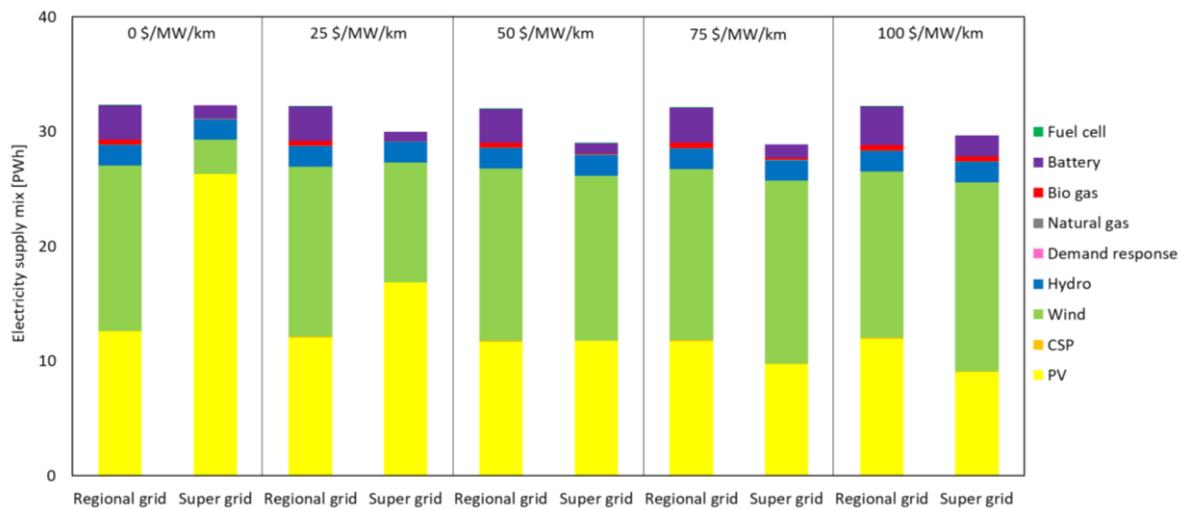

Fig. 7 Electricity supply mix for additional experiments with transmission costs ranging from 0 to 100 $/MW/km.

## 4. Discussion

The introduction of an intercontinental *Super grid* yields a 1.4%–6.7% reduction in electricity system costs. The greater economic benefits are associated with cheap transmission grids, high costs for solar PV and storage, high energy demand and a uniform discount rate across the world. Conversely, the lower end of the benefit spectrum applies to scenarios characterized by high transmission costs or high levels of flexibility, provided by cheap nuclear power or large-scale hydrogen production. In the **Base** scenario, the reduction in electricity system costs that results from the implementation of a global *Super grid* is 3.8%. In contrast, the average economic benefit achieved through *Regional grid* expansion is approximately three-fold higher, reaching 12%. Significant electricity system cost reductions are also observed when connecting India to countries in Southeast Asia (7%) and the Middle East (12%), reflecting the long-term benefits of grid expansion if these countries undergo a transition towards renewable electricity systems. Our findings also highlight that if dispatchable power generation is available and cheap to invest in (**Low nuclear** scenario), the advantages of grid expansion are constrained. This reflects the current situation in India and its neighboring nations, where immediate grid connections might yield limited benefits. However, it is essential to acknowledge the time-demanding nature of grid expansion. Establishing connections between India and countries in Southeast

Asia and the Middle East, in alignment with the first phase of the OSOWOG initiative in the coming decades, has the potential to yield substantial benefits for the future renewable electricity system in Asia. In summary, the benefits of the OSOWOG initiative's global *Super grid* expansion seem limited in comparison to *Regional grid* expansion. These findings suggest that the advantages of grid expansion for resource tapping and variation management are predominantly realized within each continent. The marginal benefit of further integrating the continents appears to decrease significantly.

Our finding regarding the benefit of the global *Super grid*, as proposed by the OSOWOG initiative, is consistent with the results of most other studies [3, 14-16, 18], where the reduction in system cost linked to allowing for a *Super grid* falls in the range of 0%–5%. In the present study, connecting the 48 subregions results in an overall system cost reduction of 16% compared to isolating the subregions (*Super grid* vs. *Isolation*). In comparison, Prol et al. [17] suggested a significantly greater benefit (74% reduction in system cost) from global electricity trading. The main difference between our study and that of Prol et al. [17] is that they only included solar PV and a generic dispatchable power generation technology, assumed zero costs for transmission grids, and did not optimize battery storage endogenously. To reveal the effect of using their method, we model one scenario with no wind power, a high storage cost, and zero transmission cost. In such a scenario, connecting the world with a *Super grid* reduces the electricity system cost by 78%, closely aligning with the cost reduction (74%) reported by Prol et al. [17]. Therefore, it is evident that the primary factor contributing to the significant difference in results is the choice of methodology. Since it is improbable that a future global energy system would rely exclusively on solar PV, and it is also unlikely that the transmission cost could reach zero, Prol et al. [17] clearly overestimate the benefits that can accrue from a *Super grid*.

We also evaluate the motivation behind the *Super grid* expansion, which relates to variation management for solar power via electricity trading. Our results show that integrating the six regions (18 time zones) always decreases investments in solar PV. Even under extreme conditions where the costs for wind, CSP and battery storage remain at the present levels, which favors the use of electricity trading to address the variable power production of the relatively cheap solar PV, we fail to observe any increased deployment of solar energy (Fig. 4). This seems to contrast with previous studies advocating for mitigating diurnal and seasonal variations in solar power through global electricity trading [21-27]. Those studies focused on the feasibility of combining diverse solar generation patterns. However, facilitating such energy transfers requires substantial expansion of long-distance transmission grids, the high costs of which negate the benefits of integrating compensatory solar power generation patterns. Therefore, from a techno-economic perspective, a global *Super grid* is not favorable for the deployment of solar power.

In this study, we apply country-specific discount rates to account for the heterogeneity of investment risks, which partly reflect how domestic social-political factors might affect the development of a *Super grid*. It is important to note that there are possibly other social and political barriers that could affect the development of a global *Super grid* [55]. One barrier relates to the risk of depending on other countries for energy supply, which is highlighted by the recent energy crisis in Europe after Russia's invasion of Ukraine. Therefore, social and political constraints, in addition to weak economic incentives, are likely to further hinder the development of a global *Super grid*. Yet, we also recognize that there are potential drivers, such as geopolitical relationships, which could propel the development of a *Super grid*. Thus, political decisions might be taken to develop a *Super grid* regardless of the economic gains. In the present study, we choose to concentrate on a pure techno-economic analysis. Our aim is to illustrate the economic baseline (system cost reduction linked to allowing for a *Super grid*) upon which any distortions would necessarily be layered. Therefore, this paper serves a useful policy purpose by characterizing the magnitude of the policy intervention required to counteract the baseline economic benefit of a global *Super grid*.

## 5. Conclusion

In this study, we assess the benefits of a global *Super grid* as proposed by the OSOWOG initiative. Our analysis involves modeling a renewable electricity system across six large regions, spanning from Australia to the US, using a capacity expansion model. We consider various assumptions regarding technology costs, heterogeneous discount rates among countries, the availability of nuclear power, hydrogen production, and uncertainties about future electricity demands.

We find that implementing a global *Super grid* leads to a 1.4%–6.7% reduction in electricity system costs. The significant impact on system costs for a renewable electricity system is contingent upon several factors, including extremely low cost for transmission grids, high costs for solar PV and storage, high energy demand, and a uniform discount rate worldwide. Overall, the benefits of a global *Super grid* are rather limited compared to *Regional grid* expansion (3.8% vs. 12% in the **Base** scenario). Notably, we do not account for additional social and political barriers linked to large-scale long-distance transmission grid expansion. Considering these barriers may render a *Super grid* even less attractive.

Furthermore, we evaluate the impact of a *Super grid* on solar power deployment. Across a wide range of cost assumptions and scenarios, we observe that allowing for a *Super grid* consistently results in decreased investments in solar power, even with transmission grids spanning 18 time zones. These results indicate that a global *Super grid*, as envisioned by OSOWOG, may not serve as an efficient tool to stimulate the deployment of solar power.

# Supplementary Materials

Chasing the eternal sun: Does a global super grid favor the deployment of solar power?

Xiaoming Kan[a,*]   Fredrik Hedenus[a]   Lina Reichenberg[a]

[a]Department of Space, Earth and Environment, Chalmers University of Technology, 41296 Gothenburg, Sweden

# Contents

---

[*] Correspondence: kanx@chalmers.se

# 1. Supplementary figures

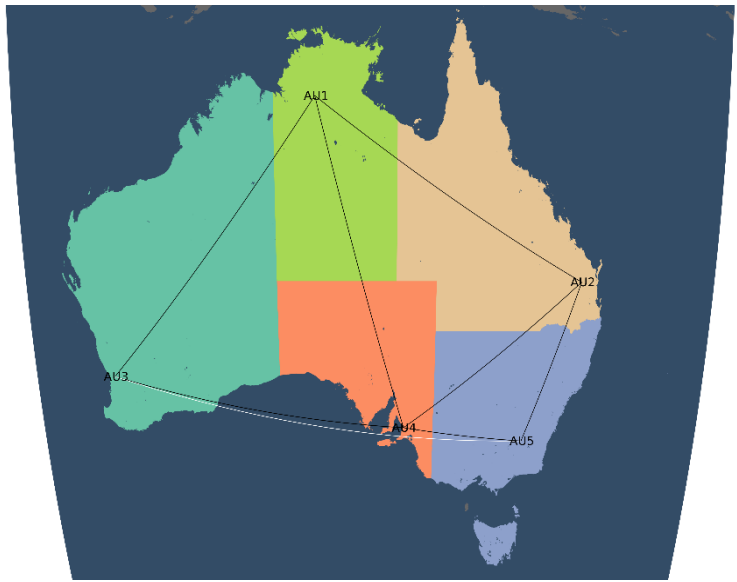

Fig. S1 Potential transmission network topology for Australia.

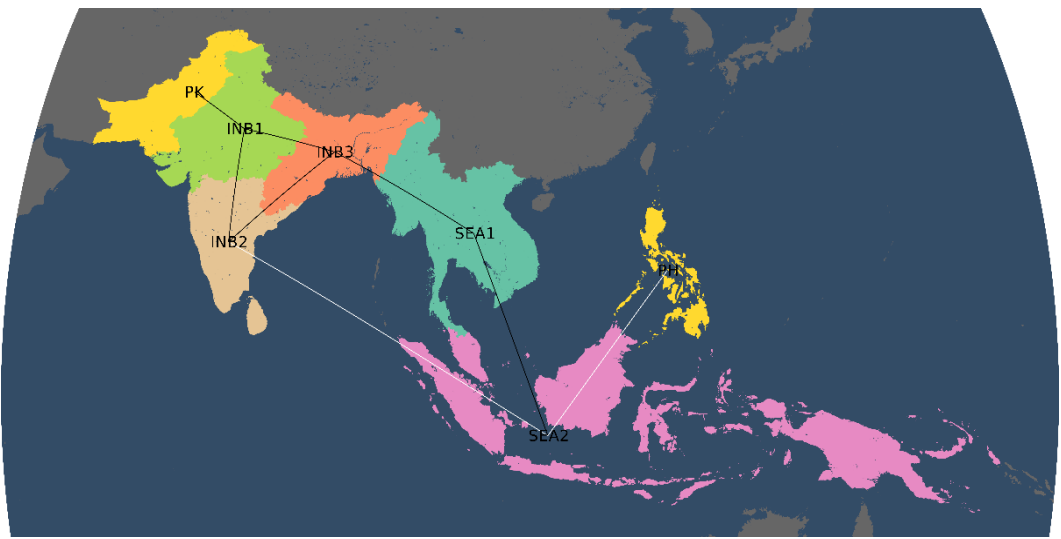

Fig. S2 Potential transmission network topology for South Asia.

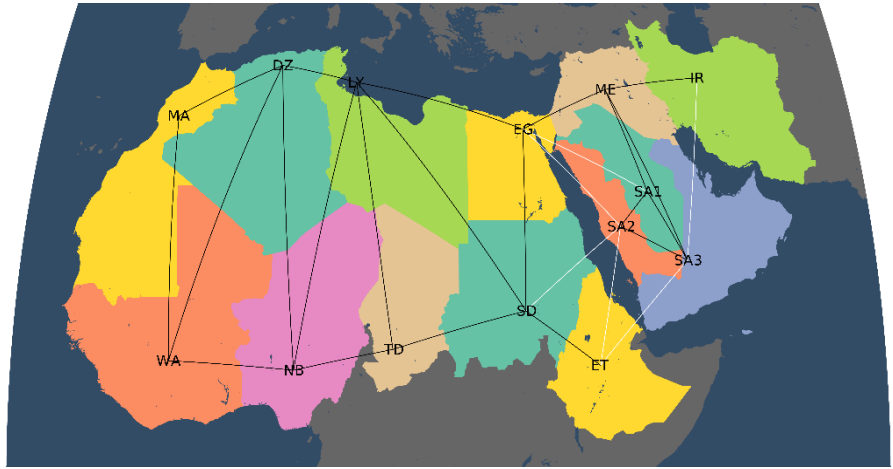

Fig. S3 Potential transmission network topology for the Middle East and North Africa.

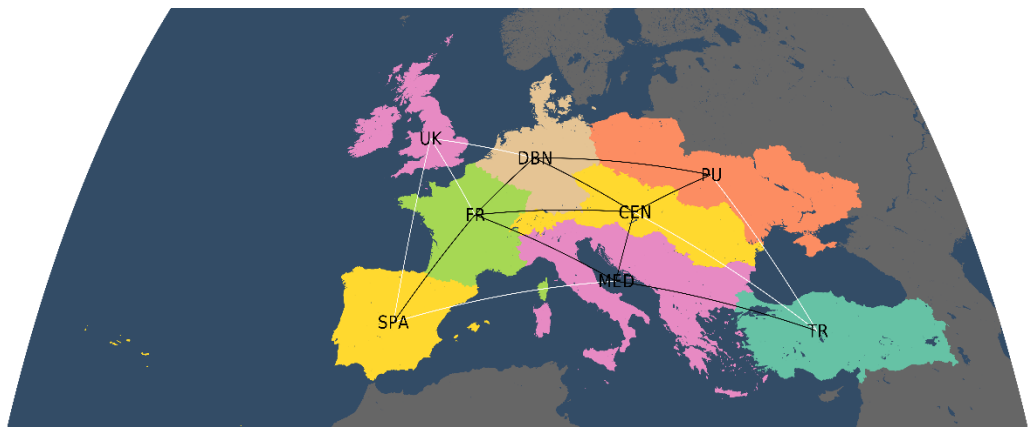

Fig. S4 Potential transmission network topology for Central and South Europe.

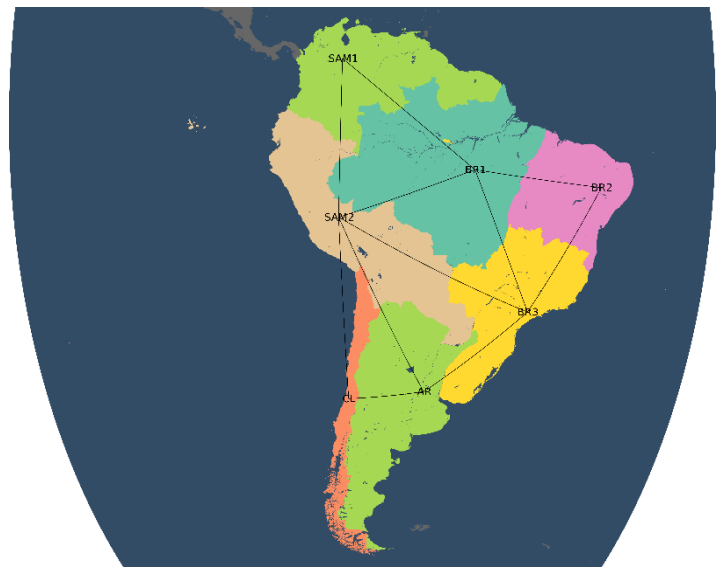

Fig. S5 Potential transmission network topology for South America.

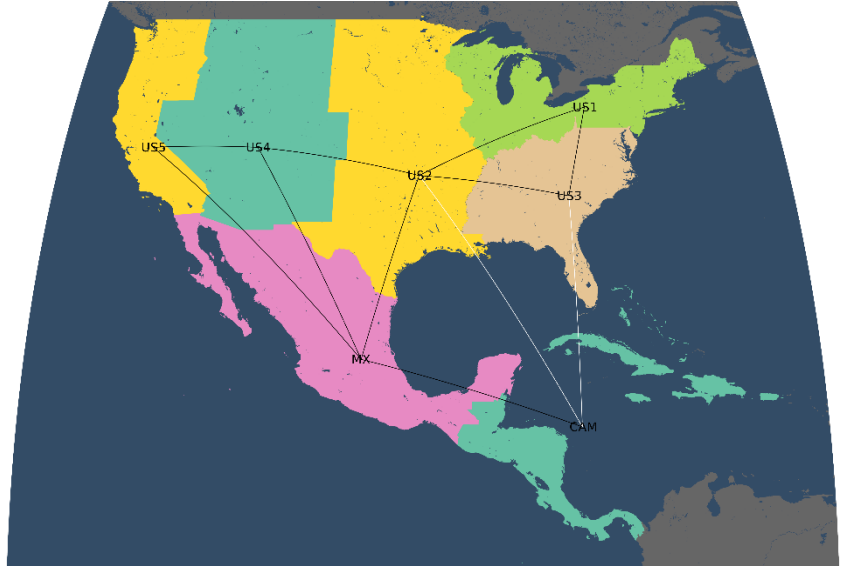

Fig. S6 Potential transmission network topology for Central and North America.

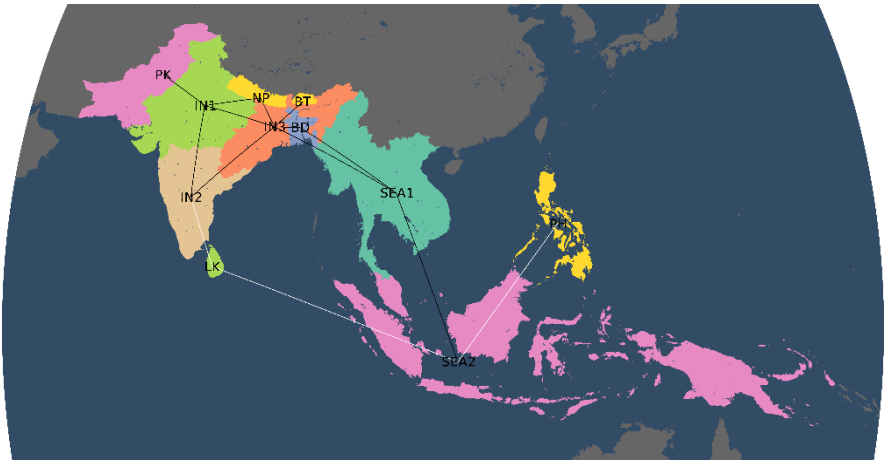

Fig. S7 Transmission grids connecting India with countries in Southeast Asia.

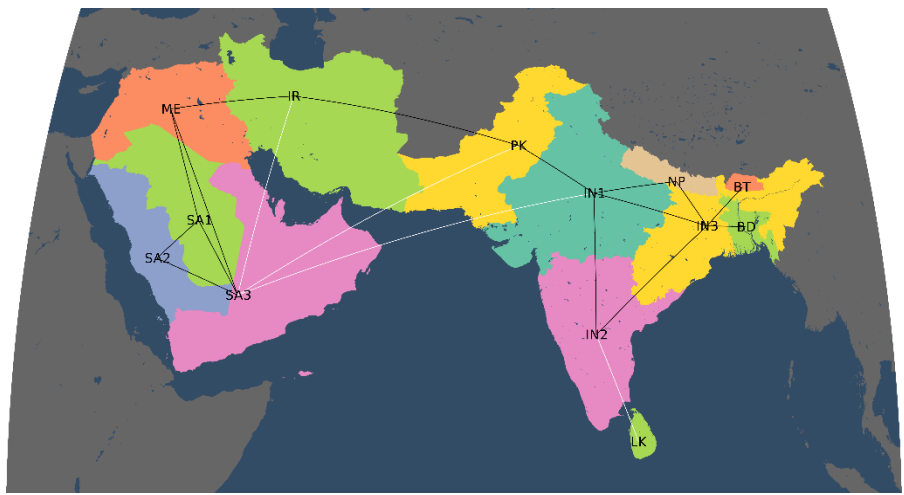

Fig. S8 Transmission grids connecting India with countries in the Middle East.

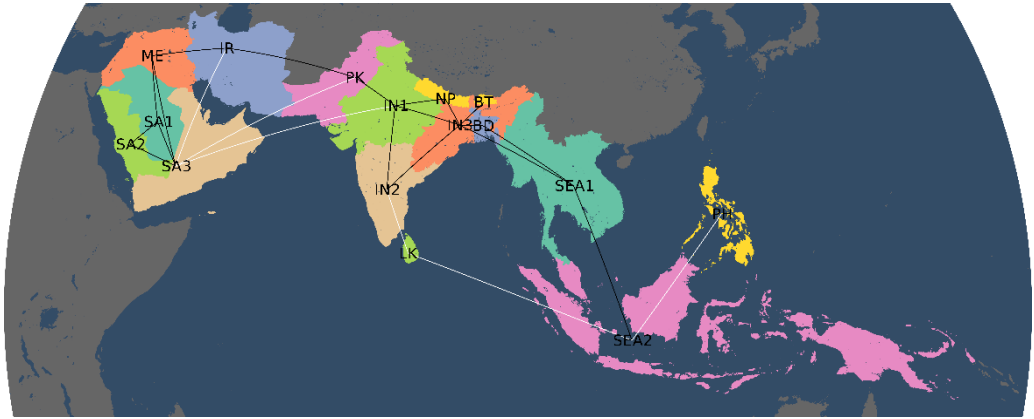

(a)

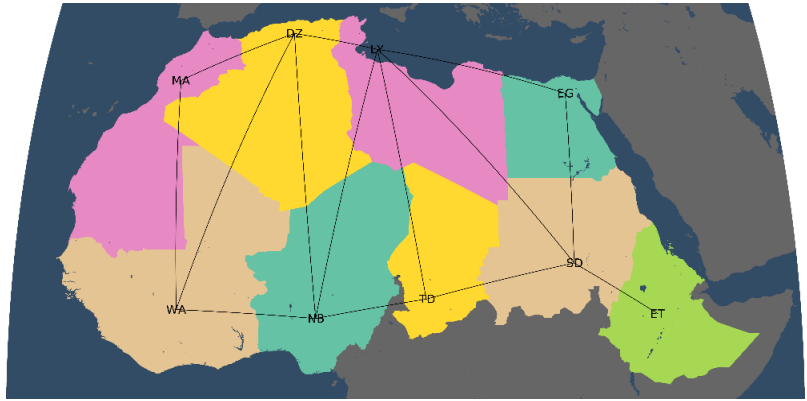

(b)

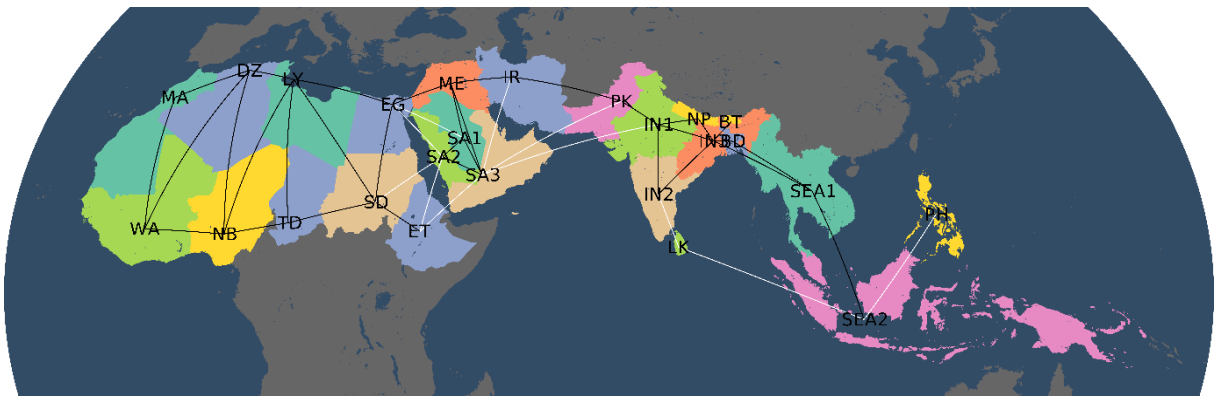

(c)

Fig. S9 Transmission grids connecting Asia and Africa. *Regional grids* in Asia (a) and Africa (b), and the connections between Asia and Africa (c).

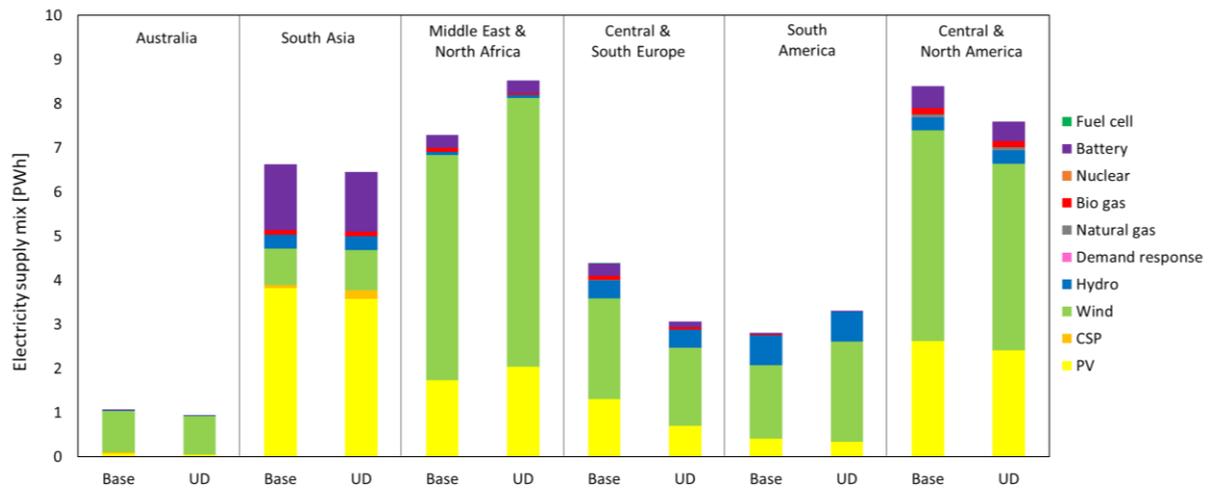

Fig. S10 Electricity supply mix in each region for the **Base** and **Uniform discount rate** (UD) scenarios.

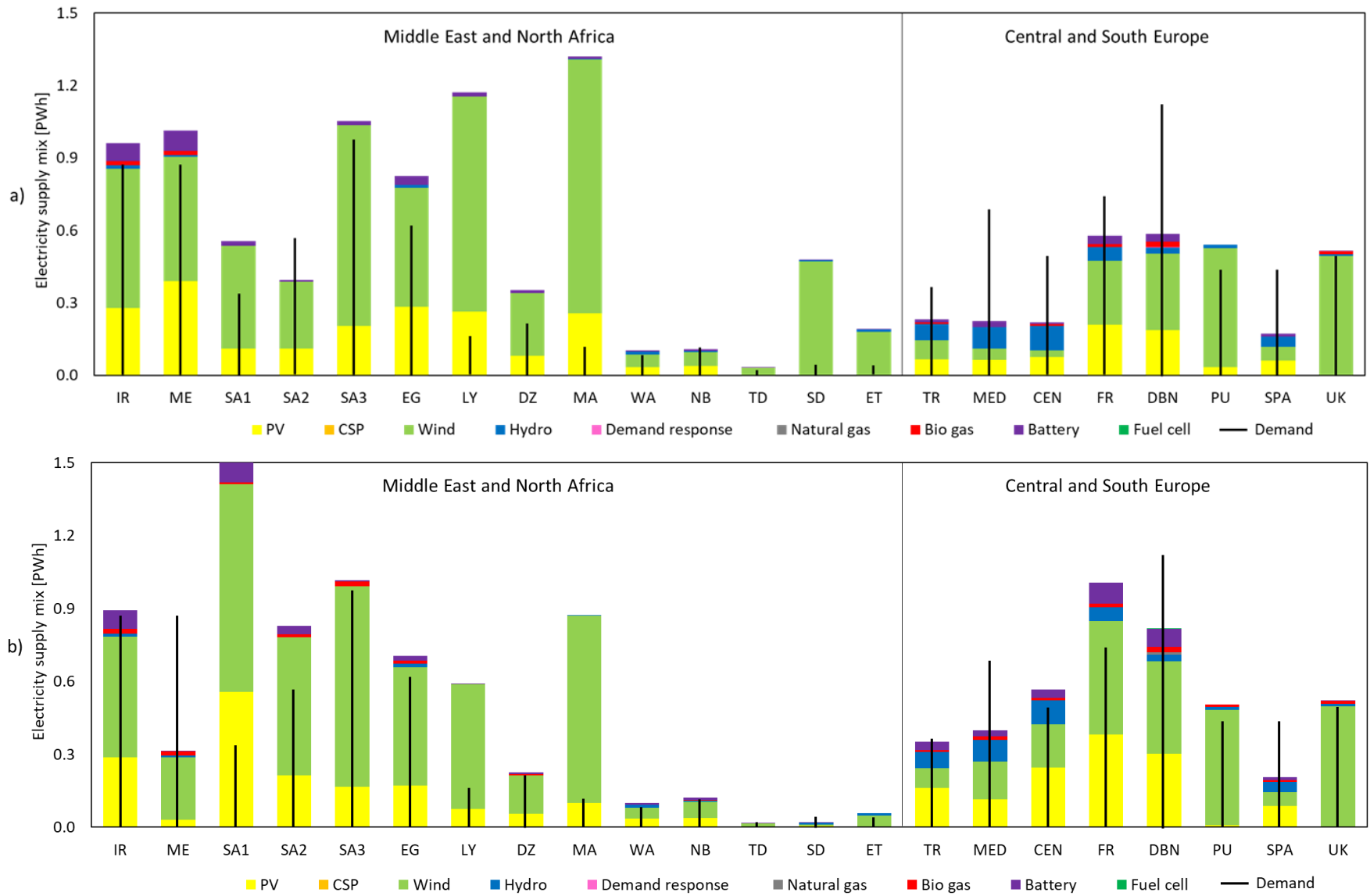

Fig. S11 Electricity supply mix for each subregion in MENA and CSE for the **Uniform discount rate** (a) and **Base** (b) scenarios.

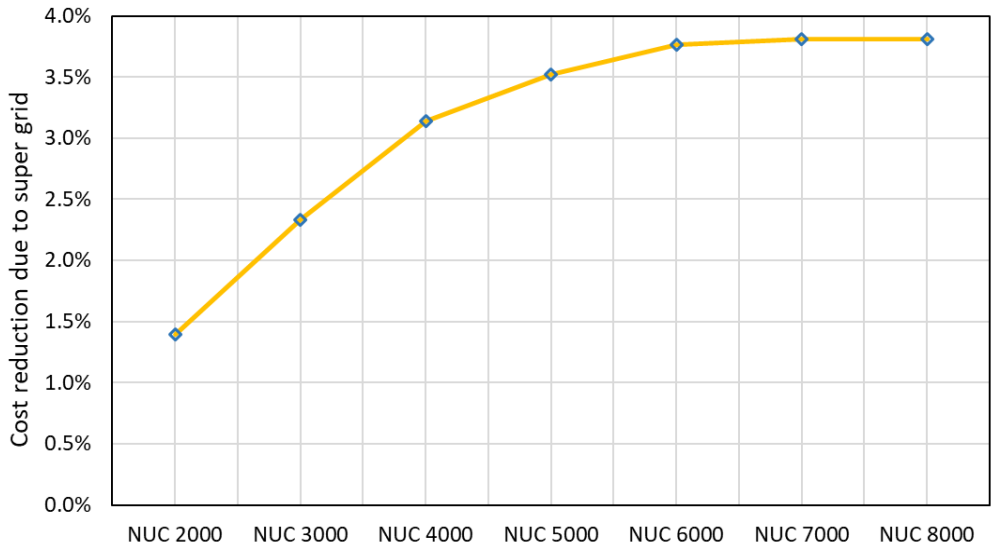

Fig. S12 Electricity system cost reductions linked to allowing for a *Super grid* under various cost assumptions for nuclear power. Specifically, nuclear power costs range from 2000 to 8000 $/kW.

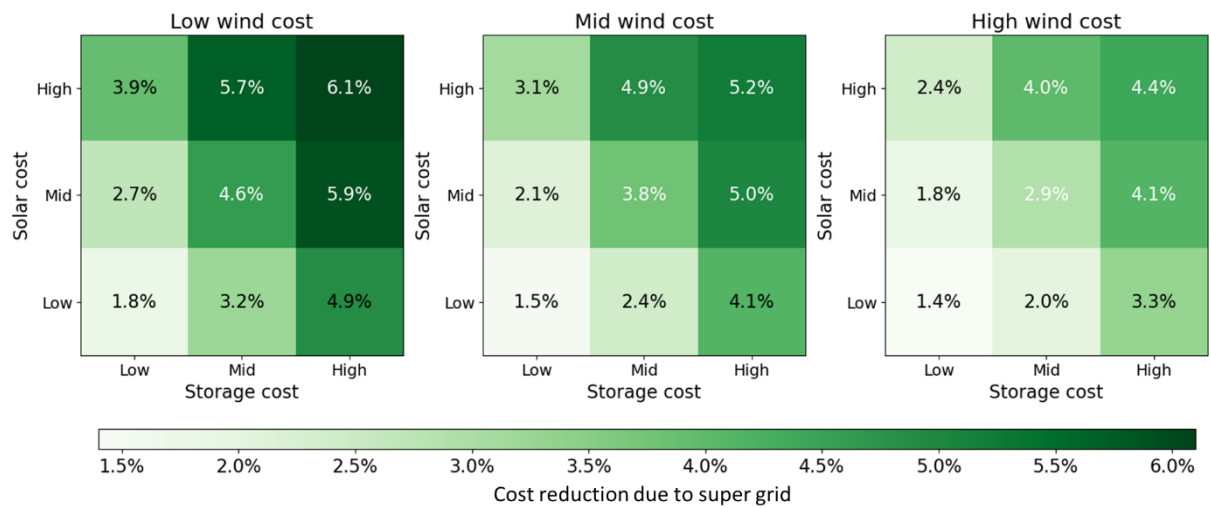

Fig. S13 Electricity system cost reductions linked to allowing for a *Super grid* under various cost assumptions for solar PV and battery storage. The initial observation highlights that the maximum impact of the *Super grid* option on reducing electricity system costs, in comparison to the *Regional grid* option, is 6.1%. This peak value for a *Super grid* emerges when both solar PV and battery storage costs are high. In this scenario, the primary generation technology that benefits from long-distance transmission (wind power) is relatively less expensive than solar power. Simultaneously, the alternative variation management strategy (storage) is comparatively more costly than transmission. Furthermore, Fig. S13 shows that scenarios with lower storage costs consistently exhibit smaller reductions in costs attributed to the *Super grid* option. This observation aligns with the intuitive notion that as the competing variation management strategy (storage) becomes more cost-effective, the importance of the variation management strategy facilitated by the *Super grid* option (intercontinental trade) diminishes. This effect is amplified if solar PV also has a low cost, as depicted along the vertical axes in Fig. S13.

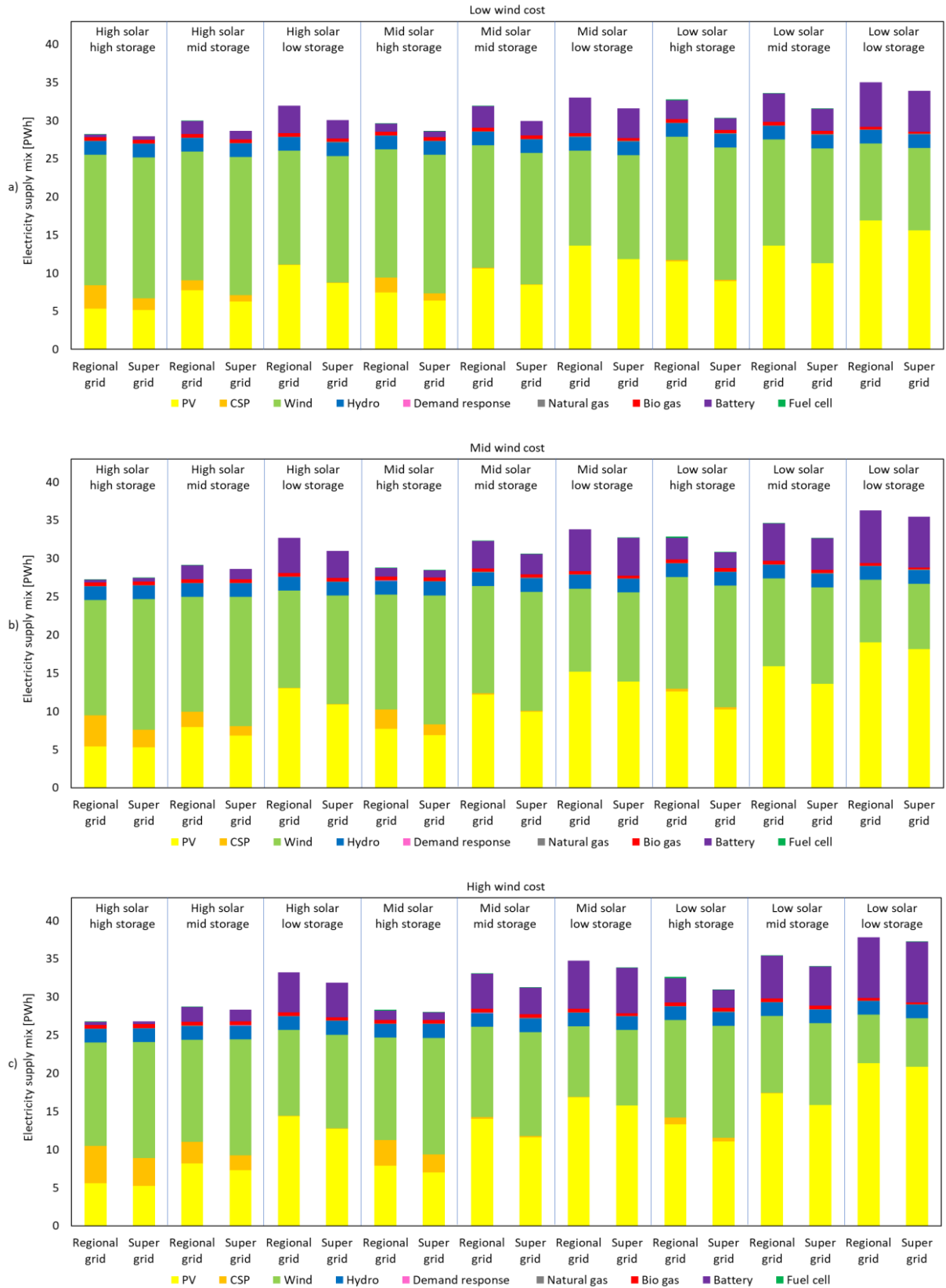

Fig. S14 Electricity supply mixes for the *Super grid* and *Regional grid* options under various cost assumptions for wind power, solar PV and battery storage.

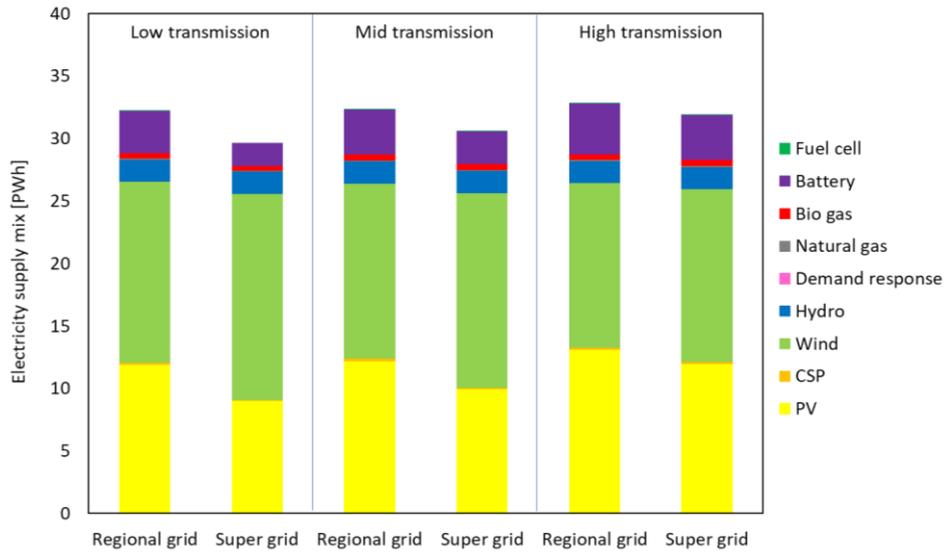

Fig. S15 Electricity supply mixes for the *Super grid* and *Regional grid* options under various cost assumptions for the transmission grid. Low transmission costs for onshore and offshore transmission grids are $150/MW/km and $200/MW/km, respectively. Mid transmission costs for onshore and offshore transmission grids are $400/MW/km and $470/MW/km, respectively, consistent with the **Base** scenario. High transmission costs for onshore and offshore transmission grids are $950/MW/km and $1000/MW/km, respectively.

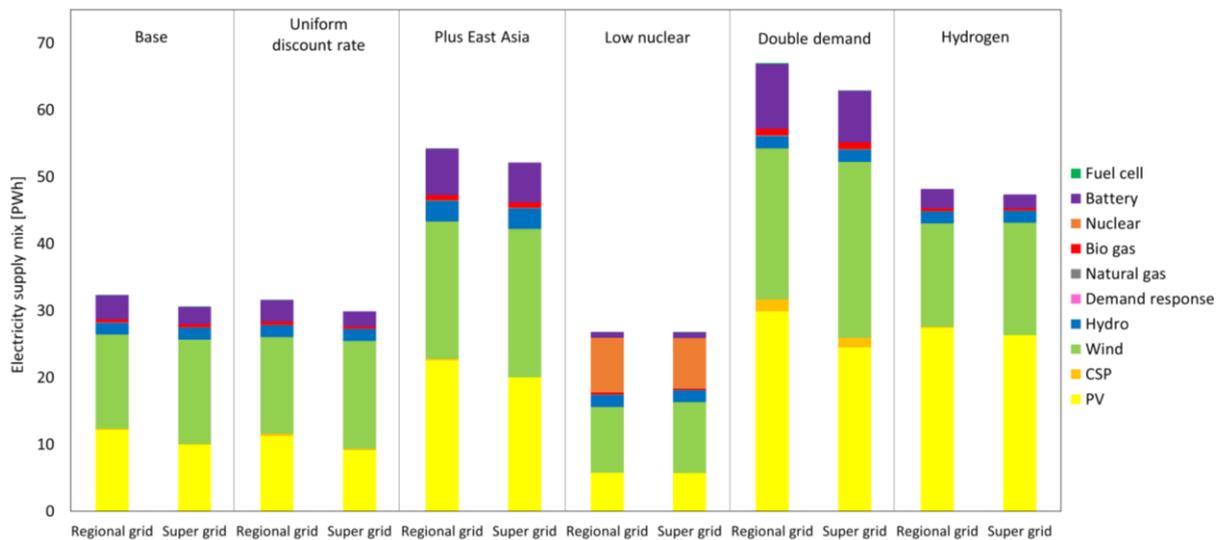

Fig. S16 Electricity supply mixes for the *Super grid* and *Regional grid* options under various scenarios. Notably, across a wide range of cost assumptions for wind, solar, battery storage and transmission grid, allowing for a *Super grid* consistently results in reduced deployment of solar power, as compared to *Regional grid* integration (Fig. S14-S15). This phenomenon holds true across various scenarios, including those with uniform discount rate, more extensive geographic areas, cheap nuclear power, higher electricity demand or hydrogen production (Fig. S16). In contrast, connecting the six continents consistently boosts the share of wind energy in the optimal electricity supply mix.

## 2. Supplementary experiments

We also investigate the benefits of connecting two neighboring regions. Integrating the adjacent regions may reduce the electricity system cost by up to 4.2% compared to isolating them, with the most significant benefit achieved when connecting MENA and CSE. Nevertheless, this benefit is considerably low compared to grid expansion within each region (12%). As for solar energy deployment, connecting neighboring regions consistently reduces the share of solar energy in the optimal electricity supply mix.

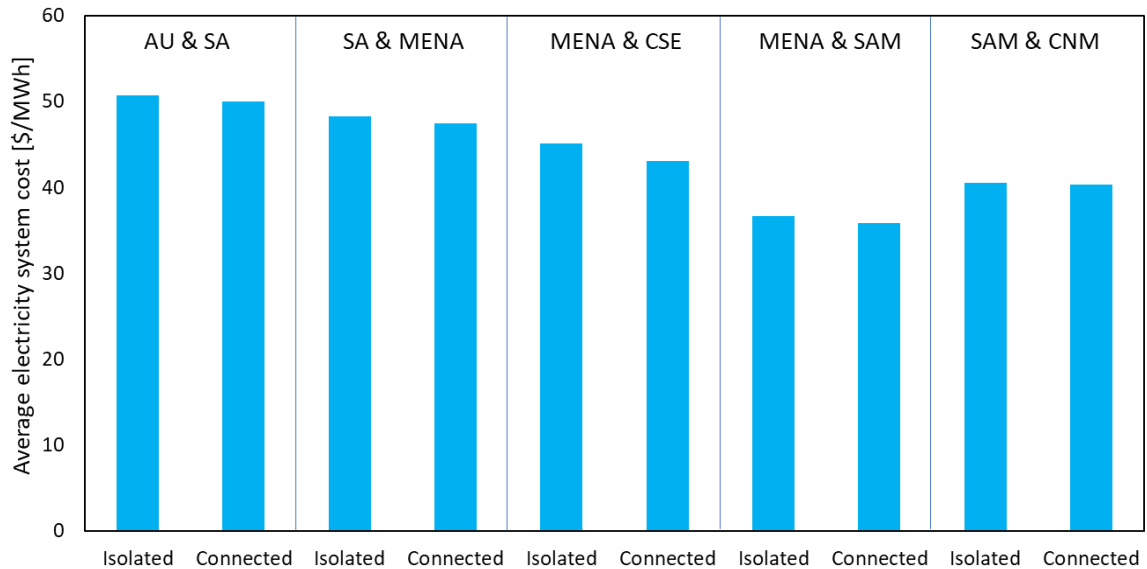

Fig. S17 Average electricity system cost before and after integrating the neighboring regions.

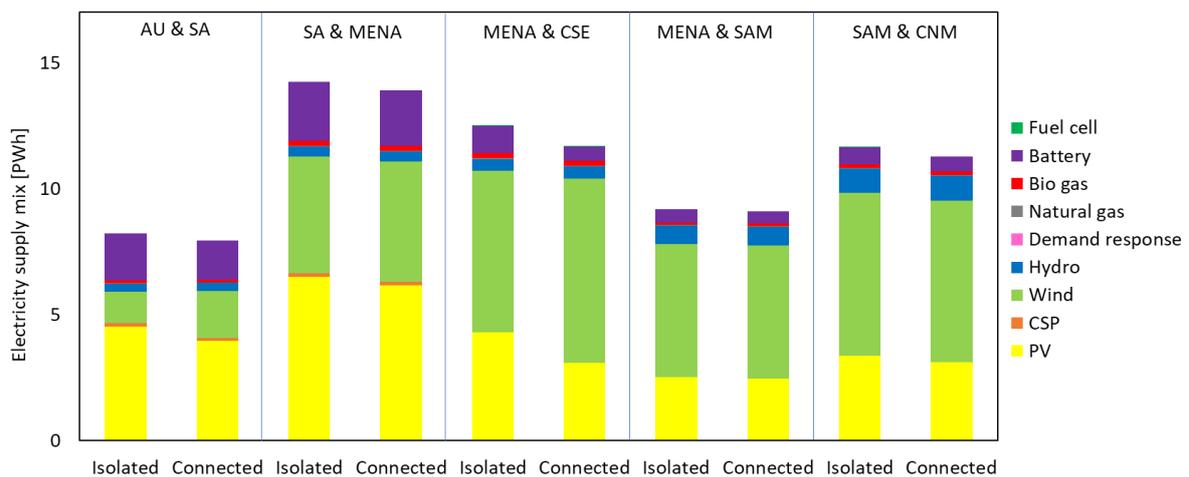

Fig. S18 Electricity supply mix before and after integrating the neighboring regions.